\documentclass[twocolumn,superscriptaddress,floatfix,prl, aps]{revtex4-1}
\usepackage{graphicx,amsfonts,amssymb,amsmath,hyperref,hypcap,enumerate,enumitem}
\usepackage[dvipsnames]{xcolor}
\usepackage{soul}
\usepackage{scalerel}
\usepackage{adjustbox,stmaryrd}
\usepackage{cancel}
\usepackage{csquotes}
\usepackage[utf8]{inputenc}
\usepackage{geometry}
\hypersetup{  colorlinks=true, linkcolor=blue, citecolor=blue, urlcolor=blue  }

\begin{document}

\title{Superconductivity in twisted transition metal dichalcogenide homobilayers}
\author{Jihang Zhu}
\email{jizhu223@gmail.com}
\affiliation{Condensed Matter Theory Center and Joint Quantum Institute, Department of Physics, University of Maryland,
College Park, Maryland 20742, USA}
\author{Yang-Zhi Chou}
\affiliation{Condensed Matter Theory Center and Joint Quantum Institute, Department of Physics, University of Maryland,
College Park, Maryland 20742, USA}
\author{Ming Xie}
\affiliation{Condensed Matter Theory Center and Joint Quantum Institute, Department of Physics, University of Maryland,
College Park, Maryland 20742, USA}
\author{Sankar Das Sarma}
\affiliation{Condensed Matter Theory Center and Joint Quantum Institute, Department of Physics, University of Maryland,
College Park, Maryland 20742, USA}

\begin{abstract}
Recently, robust superconductivity has been independently reported in twisted WSe$_2$ bilayers by two separate groups [Y. Xia et al., arXiv:2405.14784; Y. Guo et al., Nature 637, 839–845 (2025)]. 
% These observations, despite involving the same moir\'e material, were made under markedly different conditions.
In light of this, we explore the possibility of a universal superconducting pairing mechanism in twisted WSe$_2$ bilayers. Using a continuum band structure and a phenomenological boson-mediated effective electron-electron attraction, we find that intervalley intralayer pairing predominates over interlayer pairing.
Notably, despite very different experimental conditions, both twisted WSe$_2$ samples exhibit a comparable effective attraction strength for superconductivity.
This consistency in the two experiments suggests that the dominant pairing glue is likely independent of the twist angle and layer polarization, pointing to a universal underlying boson-induced pairing mechanism. We speculate on the possible mechanisms in the discussion.
\end{abstract}

% \date{\today}
{\let\newpage\relax\maketitle}
% \maketitle

% \thispagestyle{empty} % Removes page number from title page
% \newpage
% \tableofcontents
% \thispagestyle{empty} % Removes page number from table of contents
% \newpage
\setcounter{page}{1} % Starts page numbering from here

{\em Introduction.---}
\label{sec_intro}
The emergence of moir\'e superlattice in twisted transition metal dichalcogenide (TMD) homobilayers has led to the observation of a rich set of strongly correlated phases
\cite{Tang2020,Regan2020,Xu2020,Wang2020,Li2021,Ghiotto2021,Li2021a,Foutty2024,CaiJ2023a,ZengY2023a,ParkH2023,XuF2023a}, 
from correlated insulators at integer fillings to fractional Chern insulators in the absence of an external magnetic field. While the phases driven by strong electron-electron interaction manifest in the moir\'e TMD bilayers, robust observable superconductivity (SC) at dilute doping densities--associated with a few carriers per moir\'e unit cell--has remained elusive in most moir\'e TMD systems \cite{Wang2020_tWSe2}. On the other hand, both moir\'e and moir\'eless graphene multilayers have demonstrated reproducible SC with $T_c$ ranging from $20$ mK to $3$ K \cite{CaoY2018,YankowitzM2019,LuX2019,ParkJM2021,HaoZ2021,CaoY2021,OhM2021,ZhouH2021,ZhouH2022,SuR2023,ZhangY2023,HolleisL2023,LiC2024a}, indicating that SC is a generic feature in graphene-based multilayers. This difference between moir\'e TMD and moir\'e graphene systems is an important unsolved puzzle in condensed matter and material physics.

Recently, two groups have independently discovered robust SC in twisted WSe$_2$ bilayers (tWSe$_2$).
One group reported SC at a filling factor $\nu_{\rm SC}=-1$ with a twist angle of $\theta=3.65^\circ$ and a small displacement field \cite{KFMak_WSe2_2024}. The critical temperature was estimated to be $0.22$ K.
The other group observed SC around $\nu_{\rm SC}=-1.1$ with a twist angle of $\theta=5^\circ$ and a large displacement field \cite{CRDean_WSe2_2024}. The critical temperature was approximately $0.426$ K. We will refer to these two experimental samples as Sample A \cite{KFMak_WSe2_2024} and Sample B \cite{CRDean_WSe2_2024}, respectively, throughout our paper. 

%\YZC{We can modify this paragraph to be the second paragraph. I think we should take the first half reviewing their results, implications. Then, we just call the $3.65^\circ$ sample A and $5^\circ$ sample B without mentioning any technical details, such as interlayer energy difference.}

%\YZC{Add a new paragraph explaining what you do and highlighting the main findings, implications. }

In this Letter, we investigate SC in tWSe$_2$ using a realistic continuum band structure model and a phenomenological boson-mediated BCS model without assuming a specific pairing mechanism. We demonstrate the dominance of intervalley intralayer pairing with an order parameter consistent with a mixture of $s$ and $f$ waves \cite{HsuYT2021,ChouYZ2022,Akbar_SC_2024,pairing}.  Our estimates of the effective coupling constants of SC in the two experiments show very similar values, suggesting a universal pairing mechanism in tWSe$_2$ (i.e., the same bosonic glue). The possible microscopic mechanisms and the importance of the in-plane magnetic field response are also discussed. Our work represents the first systematic investigation of the SC phenomenology in tWSe$_2$ systems, paving the way for future exploration of SC in moir\'e TMD systems.

% \jihang{}{This paper is organized as follows. 
% Section~\ref{sec_bands} examines the different band structure properties of Samples A and B, resulting from their distinct twist angles and displacement fields. 
% In Sec.~\ref{sec_sc}, we use a phenomenological model to demonstrate that the electron-electron attractive interaction is dominated by intervalley intralayer pairing. Our calculations show that the critical attraction strengths for Samples A and B, based on the experimentally estimated critical temperatures, are very similar. The maximum $T_c$, evaluated as a function of $\nu$, in the two samples are nearly identical.
% In Sec.~\ref{sec_OrderParameter}, we analyze the symmetry of the supercoduncting order parameters and find that the pairing is an extended $s+f$ type.
% Finally, in Sec.~\ref{sec_inplaneB},
% we explore the effect of an in-plane magnetic field on the superconducting $T_c$.}

{\em Band Structure Model.---}
\label{sec_bands}
Our phenomenological superconducting mean-field theory is based on the low-energy continuum model of tWSe$_2$ \cite{FCW_TMDhomo_2019}, of which the valley-projected Hamiltonian is
\begin{equation}
\label{Eq_Hamil}
\mathcal{H} = 
\begin{pmatrix}
h(\pmb{q}_1) + U_1(\pmb{r}) +\frac{V_z}{2} & T(\pmb{r}) \\
T^\dagger(\pmb{r}) & h(\pmb{q}_2) + U_2(\pmb{r}) -\frac{V_z}{2}
\end{pmatrix},
\end{equation}
where $h(\pmb{q}) = -\hbar^2q^2/2m^*$ is the effective mass approximation of the valence band edge at the Dirac point $\pmb{\kappa}_l$ of layer $l=1,2$. 
$\pmb{q}_l = \pmb{k}-\pmb{\kappa}_l$ is the momentum measured from the Dirac points, and $V_z$ is the interlayer energy difference due to the displacement field. 
The layer-dependent moir\'e potential $U_l(\pmb{r})$ and the interlayer tunneling $T(\pmb{r})$ are spatially periodic with moir\'e periodicity,
\begin{gather}
U_{1,2}(\pmb{r}) = 2V\sum\limits_{j=1,3,5} \cos(\pmb{G}_j \cdot \pmb{r} \mp \psi), \\
T(\pmb{r}) = w(1+e^{-i\pmb{G}_2 \cdot \pmb{r}}+e^{-i\pmb{G}_3 \cdot \pmb{r}}). \label{Eq_DeltaT}
\end{gather}
$\pmb{G}_1 = (1,0)b_{\rm M}$, $\pmb{G}_j = \mathcal{R}_{(j-1)\pi/3} \pmb{G}_1$ are the first-shell moir\'e reciprocal lattice vectors, with $b_{\rm M} = 4\pi \theta/\sqrt{3} a_0$ as the length of moir\'e primitive reciprocal lattice vector. In the calculations throughout this paper, we use the  continuum model parameters fitted by large-scale DFT calculations \cite{TDevakul_DFT_2021}: $V=9$ meV, $\psi=128^\circ$, $w=18$ meV, lattice constant $a_0 = 3.317 \text{ \AA}$, and the effective mass $m^*=0.43m_e$ with $m_e$ being the electron mass. 

\begin{figure}[!t]
\centering
\includegraphics[width=1.0\columnwidth]{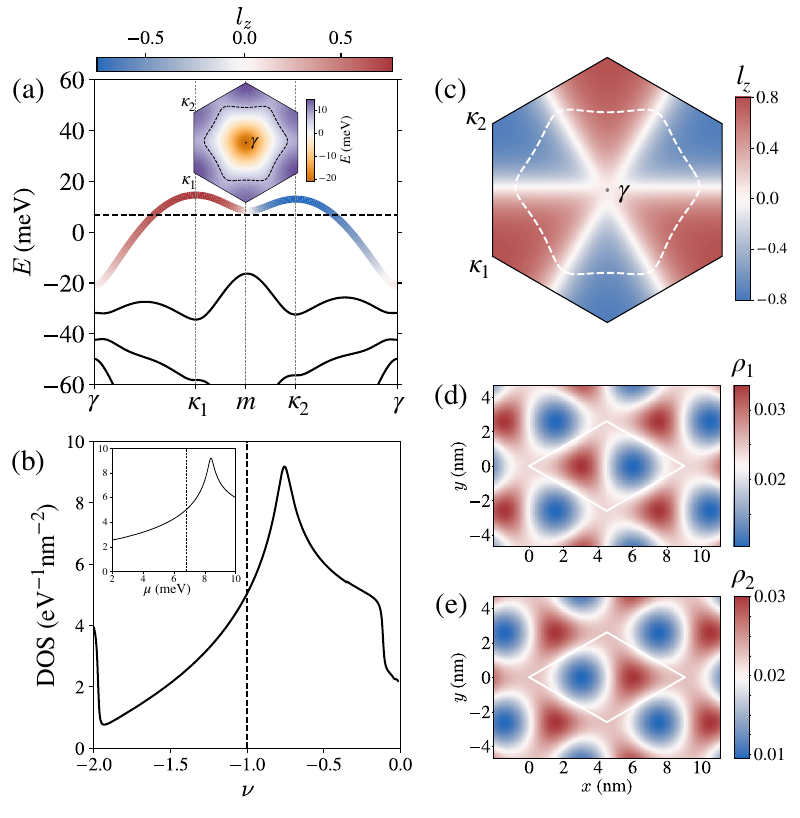}
\caption{\label{fig_spectrum_KFM} {
Band properties of Sample A: $\theta=3.65^\circ$, $V_z = 2.1$ meV and the filling factor of SC $\nu_{\rm SC} = -1$.
(a) The band structure along high-symmetry lines and the energy contour (inset) of the first moir\'e valence band. The dashed horizontal line and the dashed contour in the inset represent the chemical potential and the Fermi surface at $\nu_{\rm SC}$, respectively. The color of the first moir\'e band indicates the layer polarization $l_z$. 
(b) The DOS for $\nu \in [-2,0]$, taking into account both valleys, i.e., the filling factor $\nu$ denotes the number of electrons per moir\'e unit cell. The VHS appears around $\nu=-0.77$ and is $\sim 1.5$ meV away from the chemical potential at $\nu_{\rm SC}$ (inset).
(c) The layer polarization distribution in the first MBZ. The white dashed line outlines the Fermi surface contour at $\nu_{\rm SC}$.
(d-e) The layer-projected densities, $\rho_1$ and $\rho_2$, in real space, forming two triangular lattices centered on XM or MX local stackings. $\rho_1$ and $\rho_2$ create an effective honeycomb lattice with a small onsite energy difference.
The white hexagon marks the moir\'e unit cell. 
}}
\end{figure}

\begin{figure}
\centering
\includegraphics[width=1.0\columnwidth]{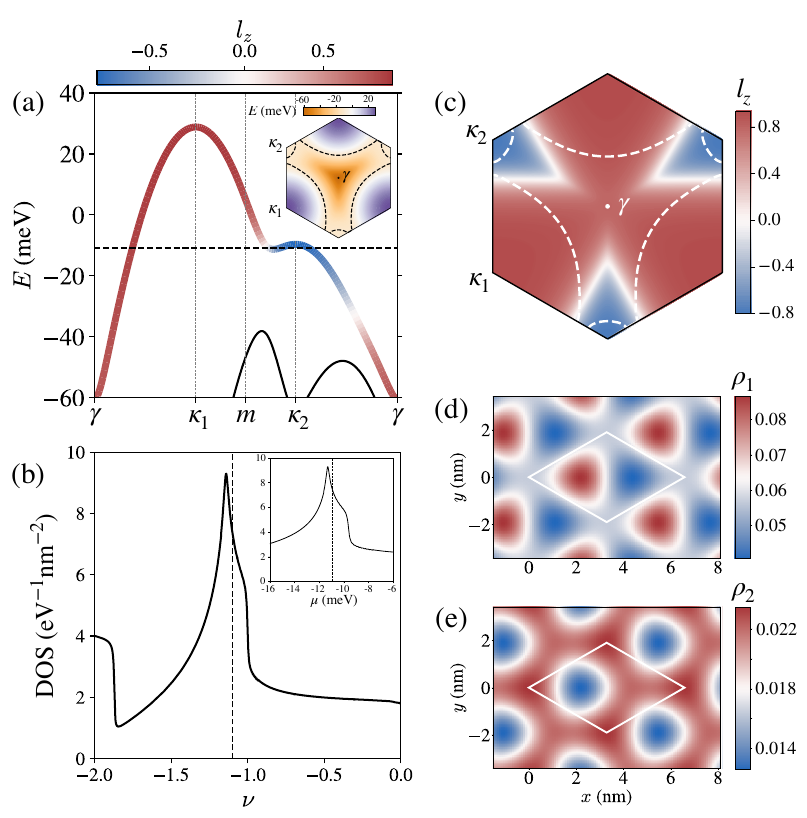}
\caption{\label{fig_spectrum_CRD} {
Band properties of Sample B: $\theta=5^\circ$, $V_z = 43.75$ meV and the filling factor of SC $\nu_{\rm SC} = -1.1$.
(a) The band structure along high-symmetry lines and the energy contour (inset) of the first moir\'e valence band.
The dashed horizontal line and the dashed contour in the inset represent the chemical potential and the Fermi surface at $\nu_{\rm SC}$, respectively.
(b) The DOS for $\nu \in [-2,0]$, showing that the VHS is near $\nu_{\rm SC}$ and less than $0.5$ meV from the chemical potential at $\nu_{\rm SC}$ (inset).
(c) The layer polarization distribution in the first MBZ.
The white dashed line outlines the Fermi surface contour at $\nu_{\rm SC}$.
(d-e) The layer-projected density distributions in real space.
  }}
\end{figure}

Figures~\ref{fig_spectrum_KFM} and \ref{fig_spectrum_CRD} show the moir\'e band structures, density of states (DOS) and layer polarization $l_z$ in momentum and real space for Samples A and B. The layer polarization operator $\hat{l}_z$ is represented by the Pauli matrix acting on the layer subspace.
In Sample A, the layer is weakly polarized due to the small displacement field, corresponding to an interlayer energy difference of approximately $2.1$ meV  \cite{KFMak_WSe2_2024, Wefollow}.
In momentum space, as shown in Fig.~\ref{fig_spectrum_KFM}(c), the wave function is localized on each layer around its Dirac point. In real space, depicted in Fig.~\ref{fig_spectrum_KFM}(d-e), the layer projected density forms two triangular lattices centered on XM or MX local stackings, creating an effective honeycomb lattice with a small onsite energy difference. The DOS plot in Fig.~\ref{fig_spectrum_KFM}(b) shows a Van Hove singularity (VHS) around $\nu=-0.77$. The SC observed at $\nu_{\rm SC}=-1$ is only $\sim 1.5$ meV away from the VHS (inset of Fig.~\ref{fig_spectrum_KFM}(b)).
% an energy scale that can be easily smeared out by the level broadening due to disorder.
In our notation, the filling factor $\nu$ represents the number of electrons per moir\'e unit cell.
In Sample B, a large displacement field induces an interlayer energy difference of $\sim 43.75$ meV  \cite{CRDean_WSe2_2024, Weapproximate}, resulting in strong layer polarization (Fig.~\ref{fig_spectrum_CRD}(c-e)).
The VHS is located at $\nu=-1.14$ (Fig.~\ref{fig_spectrum_CRD}(b)), which is less than $0.5$ meV away from the chemical potential at the filling factor $\nu_{\rm SC}=-1.1$, where SC was observed. This proximity is conducive to pairing instability. 

Despite the significant differences in band structure, layer polarization, and effective lattice model between Samples A and B, we explore the possibility that the same underlying mechanism is responsible for SC in both cases in the next section.

{\em Superconductivity.---}
\label{sec_sc}
Closely connected to the two experiments, we study the effective pairing strength in Samples A and B using a phenomenological BCS theory without assuming a specific pairing mechanism.
We will discuss the possible pairing glues, e.g., phonons and magnons, at the end of this Letter.
Given that no evidence of time-reversal symmetry (TRS) breaking was observed near the superconducting phase in either sample, we consider only the intervalley pairing that preserves TRS.
In the following part, we focus on intralayer pairing, as detailed in Supplementary Information (SI) A, the interlayer pairing is significantly weaker than intralayer pairing by a factor of five to ten. The dominance of the intralayer pairing is a direct consequence of the layer polarization patterns as shown in Figs.~\ref{fig_spectrum_KFM}(c) and \ref{fig_spectrum_CRD}(c).

The effective electron-electron attraction mediated by intralayer bosons is given by 
\begin{equation}
\label{Eq_Hatt1}
\begin{split}
H_{\rm att} = -\frac{g}{2A} 
\sum\limits_{l,\pmb{q}} \hat{n}_l(\pmb{q}) \hat{n}_l(-\pmb{q}),
\end{split}
\end{equation}
where $A$ is the system area, and
we approximate the intervalley intralayer pairing strength by a tunable static and momentum independent potential $g$. 
The density operator, $\hat{n}_l(\pmb{q})$, is defined as
% The action of the intralayer pairing is
% \begin{equation}
% \mathcal{S}_{\rm ph} = -\frac{1}{2\beta A} \sum\limits_{\nu_n,\pmb{q}} V_g(\nu_n,\pmb{q}) \sum\limits_l \hat{n}_l(\nu_n,\pmb{q}) \hat{n}_l(-\nu_n,-\pmb{q}),
% \end{equation}
% where $\nu_n$ is the Matsubara frequency, $V_g(\nu_n,\pmb{q})$ is the dynamical potential mediated by glue Bosons,
% \begin{equation}
% V_g(\nu_n,\pmb{q}) = g \frac{\omega^2_{\pmb{q}}}{\omega^2_{\pmb{q}} + \nu_n^2}.
% \end{equation}
% $\omega_{\pmb{q}}$ is the glue Bosons' dispersion. For acoustic phonon, $\omega_{\pmb{q}} = v_sq$ with $v_s$ is the sound velocity.
% We only focus on the intervalley pairing by static potential $V_g(\nu_n, \pmb{q}) \equiv g$, and using the density operator defined as
\begin{equation}
\hat{n}_l(-\pmb{q}) = \sum\limits_{\tau,\pmb{k},\pmb{G}}
\psi^\dagger_{\tau, l}(\pmb{k}+\pmb{G}) \psi_{\tau, l}(\pmb{k}+\pmb{G}-\pmb{q}),
\end{equation}
where $\tau=\pm$ is the valley index, $l=1,2$ represents the layer degree of freedom and $\pmb{k}$ is in the first moir\'e Brillouin zone (MBZ).
Thus, the effective attraction Eq.~(\ref{Eq_Hatt1}) is
\begin{equation}
\begin{split}
H_{\rm att} = -\frac{g}{A} \sum\limits_{\substack{l,\pmb{k},\pmb{k}'\\\pmb{G},\pmb{G}'}} &\psi^{\dagger}_{+,l}(\pmb{k}+\pmb{G}) \psi^{\dagger}_{-,l}(-\pmb{k}-\pmb{G}) \\
&\psi_{-,l}(-\pmb{k}'-\pmb{G}') \psi_{+,l}(\pmb{k}'+\pmb{G}').
\end{split}
\end{equation}
The factor of 2 in Eq.~(\ref{Eq_Hatt1}) is canceled by summing over valleys and keeping only the intervalley terms.
Projecting to the first moir\'e valence band of tWSe$_2$, the effective pairing Hamiltonian becomes
\begin{gather}
H_{\rm p} = -\frac{1}{A} \sum\limits_{\pmb{k},\pmb{k}'} g_{\pmb{k},\pmb{k}'} c^{\dagger}_{+}(\pmb{k}) c^{\dagger}_{-}(-\pmb{k}) c_{-}(-\pmb{k}') c_{+}(\pmb{k}'), \label{Eq_Hp} \\
g_{\pmb{k},\pmb{k}'} = g\sum\limits_{l,\pmb{G},\pmb{G}'}
|z_{+,l,\pmb{G}}(\pmb{k})|^2 |z_{+,l,\pmb{G}'}(\pmb{k}')|^2,
\label{Eq_gkkp}
\end{gather}
where $c^\dagger$ ($c$) is the quasiparticle creation (annihilation) operator in the plane-wave expansion,
\begin{equation}
c^\dagger_{\tau}(\pmb{k}) = \sum\limits_{l,\pmb{G}} z_{\tau,l,\pmb{G}}(\pmb{k}) \psi^\dagger_{\tau,l}(\pmb{k}+\pmb{G}),
\end{equation}
with $\pmb{G}$ and $\pmb{G}'$ being the moir\'e reciprocal lattice vectors. 

For $T\approx T_c$, the linearized gap equation in the BCS mean-field approximation is
\begin{equation}
\label{Eq_Delta_linearized}
\Delta_{\pmb{k}} = \frac{1}{A} \sum\limits_{\pmb{k}'} g_{\pmb{k},\pmb{k}'}
\frac{\tanh(\frac{\varepsilon_{+}(\pmb{k}')-\mu}{2k_{\rm B}T})}{2(\varepsilon_{+}(\pmb{k}')-\mu)} \Delta_{\pmb{k}'},
\end{equation}
$\varepsilon_{\tau}(\pmb{k})$ is the quasiparticle eigenenergy of the first moir\'e valence band and $\mu$ is the chemical potential.
In obtaining Eq.~(\ref{Eq_gkkp}) and (\ref{Eq_Delta_linearized}), we have used the time-reversal symmetry properties
\begin{gather}
\varepsilon_{+}(\pmb{k}) = \varepsilon_{-}(-\pmb{k}), \label{Eq_TRS_ep}\\
z_{+,l,\pmb{G}}(\pmb{k}) = z^*_{-,l,-\pmb{G}}(-\pmb{k}). \label{Eq_TRS_wf}
\end{gather}
Equation~(\ref{Eq_Delta_linearized}) can be rewritten in the matrix form:
\begin{equation}
\label{Eq_M}
\pmb{\Delta} = g\pmb{M} \pmb{\Delta}.
\end{equation}
At the critical point $T=T_c$, Eq.~(\ref{Eq_M}) has only one stable solution.
Given the electron-boson coupling strength $g$, $T_c$ is found by obtaining the largest eigenvalue of $\pmb{M}$ to be $g^{-1}$. Equivalently, for a given temperature $T_c$, the critical electron-boson coupling strength $g^*$ is the inverse of the maximum eigenvalue of $\pmb{M}$.

\begin{figure}[!t]
\centering
\includegraphics[width=1.0\columnwidth]{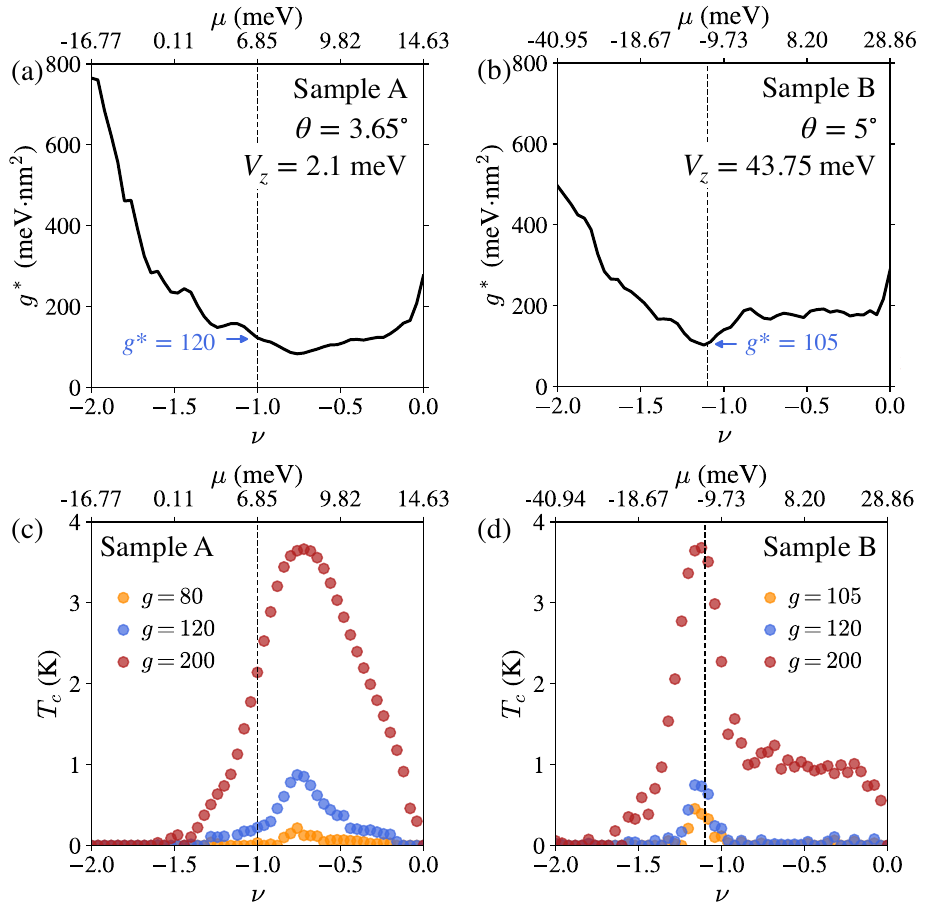}
\caption{\label{fig_gcrTc} {
(a-b) The critical electron-boson coupling strength $g^*$ determined by the experimentally estimated $T_c$ in (a) Sample A: $\theta = 3.65^\circ$, $V_z=2.1$ meV, $T_c=0.22$ K; and (b) Sample B: $\theta = 5^\circ$, $V_z=43.75$ meV, $T_c=0.426$ K.
The vertical dashed lines indicate the filling factor at which SC was observed: $\nu_{\rm SC} = -1$ in Sample A and $\nu_{\rm SC} = -1.1$ in Sample B. In (a), the minimum $g^*$ is approximately $80$ meV$\cdot$nm$^2$, and $g_{\rm SC}^* \approx 120$ meV$\cdot$nm$^2$ at $\nu_{\rm SC}=-1$.
In (b), the minimum $g^*$ is at $\nu_{\rm SC}=-1.1$ and is $g_{\rm SC}^* \approx 105$ meV$\cdot$nm$^2$. 
The chemical potential corresponding to the filling factor $\nu$ is calculated using the single-particle model and shown on the top $x$-axis.
(c-d) $T_c$ versus $\nu$ for several representative values of $g$. In both samples, $T_c$ reaches its maximum at the VHS and remains observable away from the VHS.
  }}
\end{figure}

In Fig.~\ref{fig_gcrTc}(a-b), we show $g^*$ as a function of filling factor $\nu$ for Sample A, given the experimental $T_c = 0.22$ K, and for Sample B, given the experimental $T_c = 0.426$ K.
$g^*$ reaches its minimum at the VHS for both samples, around $\nu = -0.77$ for Sample A (Fig.~\ref{fig_spectrum_KFM}(b)) and $\nu=-1.1$ for Sample B (Fig.~\ref{fig_spectrum_CRD}(b)). 
The minimum $g^*$ is approximately $80$ ($105$) meV$\cdot$nm$^2$ for Sample A (B).
At the filling factor where robust SC was observed, $g^*_{\rm SC} \approx 120$ meV$\cdot$nm$^2$ (at $\nu_{\rm SC}=-1$) for Sample A, and $g^*_{\rm SC} \approx 105$ meV$\cdot$nm$^2$ (at $\nu_{\rm SC}=-1.1$) for Sample B.
The similar values of $g^*_{\rm SC}$ in the two very different samples strongly suggest intralayer pairing with a universal bosonic glue producing SC.
We note that the interlayer pairing plays a minor role in this system, as detailed in SI B, the corresponding electron-boson coupling strength $g^*_{\rm inter}$ is at least five times larger than the intralayer $g^*$ for Sample A and ten times larger for Sample B.
Note that a larger $g^*$, calculated using the experimental $T_c$, suggests a lower likelihood of superconductivity being mediated by bosons.

In Fig~\ref{fig_gcrTc}(c-d), $T_c$ is shown as a function of $\nu$ for several representative values of $g$. In both samples, $T_c$ reaches its maximum at the VHS and remains observable away from the VHS, similar to the situation found in the graphene multilayers \cite{LothmanT2017,ChouYZ2021,ChouYZ2022a,ChouYZ2022b}. We note that non-adiabatic vertex corrections \cite{Cappelluti_SC_1996, DPhan_phononSC_2020}, which we ignore, might become important for doping densities very close to VHS. This is an interesting subject for future work.
% Using the same $g$, for example $g=120$ meV$\cdot$nm$^2$ (blue dots in Fig~\ref{fig_gcrTc}(c-d)) or $g=200$ meV$\cdot$nm$^2$ (red dots in Fig~\ref{fig_gcrTc}(c-d)), both samples exhibit the same maximum $T_c$, indicating a common origin of SC in these two samples.
To further explore the dependence of $g^*$ and $T_c$ on the displacement field, which is a common experimental tuning parameter, we show phase diagrams of $g^*$ and $T_c$ in SI B for twist angles $\theta=3.65^\circ$ and $\theta=5^\circ$. In both cases, the minimum $g^*$ and maximum $T_c$ track the VHS.

% Our phenomenological theory ignores band renormalization due to Coulomb interactions. However, our qualitative estimation of $g^*$ remains valid since strong intraband screening near the VHS weakens the coulomb interaction. A detailed investigation of the Coulomb interaction will be addressed in future studies. \YZC{I have discussed the Coulomb interaction in the discussion. Maybe suppress this part?}

%\YZC{Actually, the $g^*$ here has incorporated Coulomb in some sense. The subtlety is that the projected interaction takes a slightly different form, but the differences are going away in the layer polarized limit. Remove the disucssion about the large bnandwidth. We can just say that the proximity to VHS generically results in a strong intraband screening, suppressing the Coulomb interaction.}

For $T < T_c$, the order parameter $\Delta_{\pmb{k}}$ is solved self-consistently,
\begin{equation}
\begin{split}
\Delta_{\pmb{k}}
&= \frac{1}{A} \sum\limits_{\pmb{k}'} g_{\pmb{k},\pmb{k}'}
\frac{\tanh(\frac{\sqrt{\xi^2_{\pmb{k}'}+|\Delta_{\pmb{k}'}|^2}}{2k_{\rm B}T})}{2\sqrt{\xi^2_{\pmb{k}'}+|\Delta_{\pmb{k}'}|^2}} \Delta_{\pmb{k}'},
\end{split}
\end{equation}
where $\xi_{\pmb{k}} = \varepsilon_+(\pmb{k}) - \mu$. At $T=0$, the hyperbolic tangent term simplifies to 1.
% The even and odd parity components of $\Delta_{\pmb{k}}$ are
% \begin{gather}
% \Delta^{\rm even}_{\pmb{k}} = \frac{1}{2} \big(\Delta_{\pmb{k}} + \Delta_{-\pmb{k}} \big), \\
% \Delta^{\rm odd}_{\pmb{k}} = \frac{1}{2} \big(\Delta_{\pmb{k}} - \Delta_{-\pmb{k}} \big).
% \end{gather}

\begin{figure}[!t]
\centering
\includegraphics[width=1.0\columnwidth]{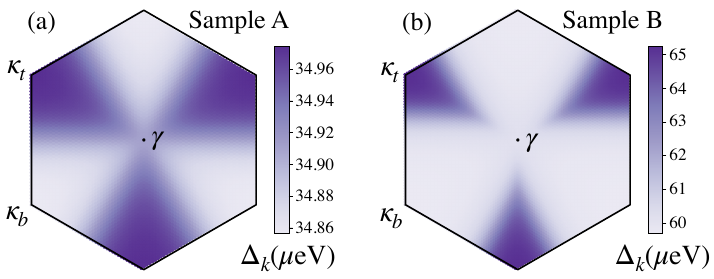}
\caption{\label{fig_Deltak}
The order parameter $\Delta_{\pmb{k}}$ for (a) Sample A and (b) Sample B. In (a), $\bar{\Delta}_k/k_BT_c = 1.84$ and in (b) $\bar{\Delta}_k/k_BT_c = 1.66$, where $\bar{\Delta}_k$ is the $k$-space average of $\Delta_k$. %\YZC{Is the order parameter peaked at the Fermi surface? } \jihang{}{The order parameter is peaked at the maximum top layer polarization.}
}
\end{figure}

The $k$-space distributions of the order parameters for Samples A and B, calculated at $T=0$, are shown in Fig.~\ref{fig_Deltak}.
Since we consider only intralayer pairing, the symmetry of $\Delta_{\pmb{k}}$ closely matches that of the layer polarization shown in Fig.~\ref{fig_spectrum_KFM}(c) and Fig.~\ref{fig_spectrum_CRD}(c), indicating a mixture of $s$ and $f$ waves \cite{Zegrodnik_SC_tWSe2_2023, Akbar_SC_2024}. 
In Sample A, $\bar{\Delta}_k/k_BT_c = 1.84$, and $\bar{\Delta}_k/k_BT_c = 1.66$ in Sample B, where $\bar{\Delta}_k$ is the $k$-space average of $\Delta_k$.
Both ratios are close to the BCS mean-field value of $1.75$. 

{\em Discussion.---}
We develop a BCS theory for the recently observed SC in tWSe$_2$. We establish that the SC likely arises from the same bosonic glue in two different experiments with different twist angles, displacement fields, and doping levels. The dominant pairing is intervalley intralayer interaction with an $s+f$ order parameter symmetry, and the maximum $T_c$ is not far from the VHS. Take the acoustic phonon as an example \cite{WuF2019,WuF2020,LewandowskiC2021,ChouYZ2021,ChouYZ2022a,ChouYZ2022b,BostromEV2023} for the bosonic glue, the calculated $g^*$ is related to the deformation potential by $D = v_s\sqrt{g^* \rho_m}$. Using the mass density $\rho_m = 6.2 \times 10^{-7}$ g/cm$^2$ and sound velocity $v_s=3.3 \times 10^5$ cm/s of monolayer WSe$_2$ \cite{Jin_TMDphonon_2014},  $g^*_{\rm SC}$ of Samples A and B in Fig.~\ref{fig_gcrTc}(a-b)
correspond to deformation potentials of $7.1$ eV and $6.7$ eV, respectively, which are in the same order of magnitude as $D=3.2$ eV for monolayer WSe$_2$ estimated from previous density functional theory (DFT) calculations \cite{Jin_TMDphonon_2014}. The quantitative extraction of the deformation potential is difficult and DFT estimates are often much smaller than experimental values, a common occurrence in semiconductors.

It is also possible that the SC observed in both samples is mediated by magnons, as suggested by the close proximity of the SC to the correlated state, very likely of an antiferromagnetic order, and the $T_c$ that peaks near the phase boundary between these states.
Understanding the magnetic order, potentially through a tight-binding model \cite{Crepel_TMD_TBmodel_2024}, is essential for characterizing the nature of spin fluctuations and the resulting superconducting state.
At present, the underlying mechanism of SC remains an unresolved and intriguing question that future experimental and theoretical work should address.
Additionally, near a VHS, repulsive interactions may lead to competing orders such as (pseudo)spin density waves, which could influence the SC phase. These competing orders may either coexist with or suppress SC under certain conditions, depending on factors like dielectric constant, doping density and displacement field. A thorough exploration of the interplay between SC and these competing phases remains an open problem and requires further investigation.

%\YZC{ We should quickly comment on the possibility of magnon-induced SC as both experiments find SC nearby some insulating states. Then, say this is an open question for future work. Finally, reiterate that our work suggests that SC in Sample A and B may come from the same universal pairing mechanism, which should be determined by future experimental and theoretical work.}

% \jihang{}{In this Letter, we investigate the possibility of a universal SC pairing mechanism in the two tWSe$_2$ samples using a phenomenological BCS model.
% Our approach is based on the validity of the Migdal theorem, which asserts that the velocity of collective bosons is much smaller than that of electrons near the Fermi surface \cite{Cappelluti_SC_1996, DPhan_phononSC_2020}.
% } \YZC{My point was to mention a potential outstanding issue as we use BCS theory in the regime where Migdal theorem might fail. This does not look good at here as we do not have a long discussion. I will incorporate this to the main text. Remove this part.} 

%\YZC{Maybe we can relocate this discussion to the end of the paper. We also want to check whether there are multiple acoustic phonon modes. In that case, the estimate of $g$ is different. I think we should not convert the $g$ to $D$ as there might be multiple pairing glues.}

Next, we comment on the role of Coulomb interactions. Coulomb interactions may result in correlated states \cite{KimS2024} that could preempt SC predicted by our phenomenological BCS theory. The Coulomb repulsion in the Cooper channel is effectively captured by our theory through the effective coupling constant $g$. However, the static intraband screening depends on the density of states, likely resulting in a more drastic change in $T_c$ (than our current phenomenological BCS approach) close to VHS \cite{ChouYZ2022a}. Microscopic calculations incorporating the frequency-dependent pairing and Coulomb interactions are necessary for a complete quantitative understanding of the underlying SC mechanism in tWSe$_2$ \cite{KFMak_WSe2_2024,CRDean_WSe2_2024}, but our current work establishes a bosonic glue to be the likely microscopic mechanism.
Moreover, the band renormalization effects due to Coulomb interactions, which we have ignored, may quantitatively modify the band structure used in this Letter. Regarding the single-particle band structure, a more accurate description which utilizes neural networks \cite{Zhang_MLDFT_2024} suggest that for large twist angles ($\theta \gtrsim 5^\circ$), the topmost moir\'e valence bands of tWSe$_2$ likely originate from the $\Gamma$-valley, rather than the $K$-valley as assumed in our model. The potential new physics associated with the low-energy bands from the $\Gamma$-valley remains an open question for future investigation.

Finally, we discuss the response to an in-plane magnetic field, which has been an important tool to discern the properties of SC in graphene-based materials \cite{CaoY2021,ZhouH2021,ZhouH2022,SuR2023,ZhangY2023,HolleisL2023,LiC2024a}. In WSe$_2$, the Zeeman effect can be ignored due to large Ising spin-orbit coupling. However, the orbital effect can still be nontrivial as the separation between two WSe$_2$ layers is not small. We find that SC is suppressed by an in-plane magnetic field of a few Teslas because the nesting of intervalley pairing requires $\varepsilon_+(\pmb{k})=\varepsilon_-(-\pmb{k})$, which is easily violated in the presence of a magnetic field. Additionally, an in-plane magnetic field may influence nearby correlated states; for example, in an antiferromagnetic state, spin fluctuation can be reduced by an applied magnetic field, weakening fluctuation-induced pairing. Systematic investigations along these lines are essential for a complete understanding of SC in tWSe$_2$.

%\jihang{}{The response of SC to an in-plane magnetic field $B_{\parallel}$ is likely weak due to the relatively large bandwidth in tWSe$_2$. We estimate that the superconductivity is suppressed by an in-plane magnetic field of a few Teslas.} 

% For any realistic magnetic field strength, the momentum shift $\delta_k$ as a result of an $B_{\parallel}$ is much smaller than the moir\'e primitive reciprocal lattice vector $b_{\rm M}$: given $B_{\parallel} = 1$ T, $\delta_k/b_{\rm M} \sim 0.0003$. 

%\YZC{I don't think this is the way we want to frame it. Your preliminary results do show suppression of $T_c$ upon a few Tesla. The change is $k$ is not really important. I think the change in $E_+(k)-E_-(k)$ is the quantity that actually matters. We can say a few ongoing directions with the in-plane magnetic field. Then, you can say we estimate that the superconductivity might be suppressed by an in-plane magnetic field with few Tesla without providing much details. We can investigate this issue later.}

{\em Acknowledgments.---}
We acknowledge the valuable discussion with Cory R. Dean.
Y.-Z. C. and J. Z. also thank Yuting Tan, Yi Huang and Jay D. Sau for useful conversations. 
This work is supported by the Laboratory for Physical Sciences.

\newpage
\newpage
%%%%%%%%%% Prefix a "S" to all equations, figures, tables and reset the counter %%%%%%%%%%
\setcounter{equation}{0}
\setcounter{figure}{0}
\setcounter{table}{0}
\setcounter{page}{1}
\makeatletter
\renewcommand{\theequation}{S\arabic{equation}}
\renewcommand{\thefigure}{S\arabic{figure}}
\renewcommand{\thetable}{S\arabic{table}}
%\renewcommand{\bibnumfmt}[1]{[S#1]}
%\renewcommand{\citenumfont}[1]{S#1}
%%%%%%%%%% Prefix a "S" to all equations, figures, tables and reset the counter %%%%%%%%%%
\onecolumngrid

\section{Supplementary Information}
\section{A. The Interlayer pairing}
\label{interlayer_pairing}
\begin{figure}[!b]
\centering
\includegraphics[width=0.65\columnwidth]{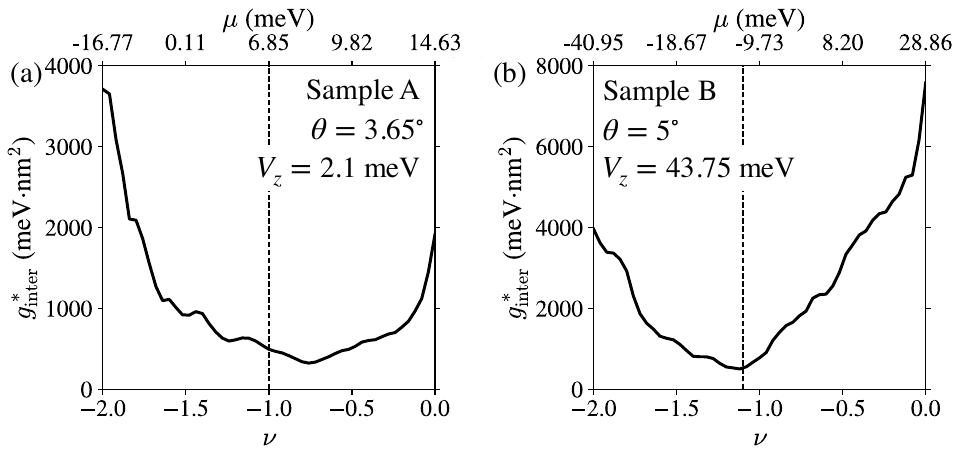}
\caption{\label{fig_ginter} {
The critical interlayer pairing strength, $g^*_{\rm inter}$, calculated using experimentally estimated $T_c$ in Samples A and B.
(a) In Sample A, which is weakly layer polarized, $g^*_{\rm inter}$ as a function of $\nu$ follows the trend of intralayer $g^*$ in Fig.~\ref{fig_gcrTc}(a), but is scaled up by a factor of $\sim 5$.
(b) In Sample B, which is strongly layer polarized due to a large displacement field, $g^*_{\rm inter}$ is at least ten times larger than the intralayer $g^*$ in Fig.~\ref{fig_gcrTc}(b). $g^*_{\rm inter}$ is maximized at the band edges, i.e., $\nu \rightarrow 0$ or $-2$, where the layer is strongly polarized (Fig.~\ref{fig_spectrum_CRD}). 
Both intralayer ($g^*$) and interlayer ($g^*_{\rm inter}$) coupling strengths are minimized near the VHS in both samples.
  }}
\end{figure}

\begin{figure}[!t]
\centering
\includegraphics[width=1.0\columnwidth]{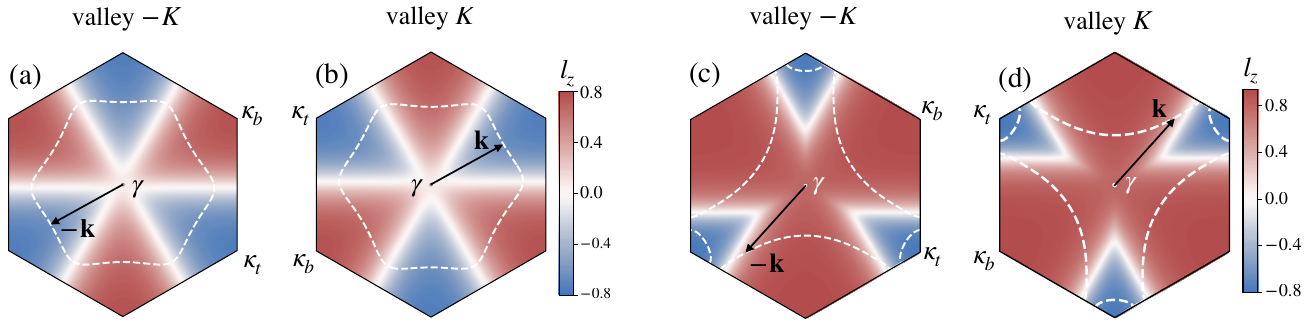}
\caption{\label{fig1_response} {
Layer polarization in $k$-space projected to valley $K$ and valley $-K$ for (a-b) Sample A and (c-d) Sample B.
Intervalley Cooper pairs form between momentum $\pmb{k}$ in valley $K$ and $-\pmb{k}$ in valley $-K$. Consequently, only a small fraction of $k$-space, where the layer polarization vanishes, contributes to interlayer pairing.
  }}
\end{figure}

Similar to the derivation of intralayer pairing in the main text,
the effective electron-electron attraction mediated by interlayer pairing is
\begin{equation}
\begin{split}
H_{\rm att}^{\rm inter} &= -\frac{1}{2 A} \sum\limits_{\pmb{q}} V_g(\pmb{q}) \sum\limits_l \hat{n}_l(\pmb{q}) \hat{n}_{\bar{l}}(-\pmb{q}) \\
&= -\frac{g}{A} \sum\limits_{\pmb{k},\pmb{k}',l} \psi^{\dagger}_{+,l}(\pmb{k}) \psi^{\dagger}_{-,\bar{l}}(-\pmb{k}) \psi_{-,\bar{l}}(-\pmb{k}') \psi_{+,l}(\pmb{k}'),
\end{split}
\end{equation}
where $\bar{l}$ represents the opposite layer of $l$.
The corresponding pairing Hamiltonian and effective coupling are
\begin{gather}
H_{\rm p}^{\rm inter} = -\frac{1}{A} \sum\limits_{\pmb{k},\pmb{k}'} g^{\rm inter}_{\pmb{k},\pmb{k}'} c^{\dagger}_{+}(\pmb{k}) c^{\dagger}_{-}(-\pmb{k}) c_{-}(-\pmb{k}') c_{+}(\pmb{k}'), \\
g^{\rm inter}_{\pmb{k},\pmb{k}'} = g\sum\limits_{l,\pmb{G},\pmb{G}'}
z^*_{+,l\pmb{G}}(\pmb{k}) z_{+,\bar{l}\pmb{G}}(\pmb{k}) z^*_{+,\bar{l}\pmb{G}'}(\pmb{k}') z_{+,l\pmb{G}'}(\pmb{k}').
\end{gather}
Similar to Fig.~\ref{fig_gcrTc}(a-b) in the main text, the critical electron-boson coupling strength $g^*_{\rm inter}$ for interlayer pairing is shown in Fig.~\ref{fig_ginter}. For Sample A, $g^*_{\rm inter}$ as a function of $\nu$ exhibits a similar trend to the $g^*$ obtained from intralayer pairing (Fig.~\ref{fig_gcrTc}(a)), but is scaled up by a factor of $\sim 5$. This is because the layer-projected wave function is localized in separate regions in $k$-space, as shown in Fig.~\ref{fig_spectrum_KFM}(c), and the layer polarization has a weak dependence on $\nu$ due to the small displacement field.
For Sample B, however, $g^*_{\rm inter}$ is much larger than the intralayer pairing $g^*$ (Fig.~\ref{fig_gcrTc}(b)) at small filling $|\nu| \lesssim 0.4$, the interlayer pairing strength is at least ten times weaker than that of the intralayer pairing. The interlayer pairing is much weaker for larger layer polarization, this can be understood by the layer polarization distribution in $k$-space. As shown in Fig.~\ref{fig1_response}, layer polarization is projected to valley $K$ and valley $-K$, and intervalley Cooper pairs form between momentum $\pmb{k}$ in valley $K$ and $-\pmb{k}$ in valley $-K$. Consequently, only a small fraction of $k$-space, where the layer polarization vanishes, contributes to interlayer pairing.
With increasing displacement field, this fraction becomes even smaller, as depicted in 
Fig.~\ref{fig1_response}(c-d), leading to a further increase in $g^*_{\rm inter}$ in Fig.~\ref{fig_ginter}(b).

\section{B. Superconductivity phase diagram}
\label{sec_PhaseDiagram}
\begin{figure}[!t]
\centering
\includegraphics[width=0.7\columnwidth]{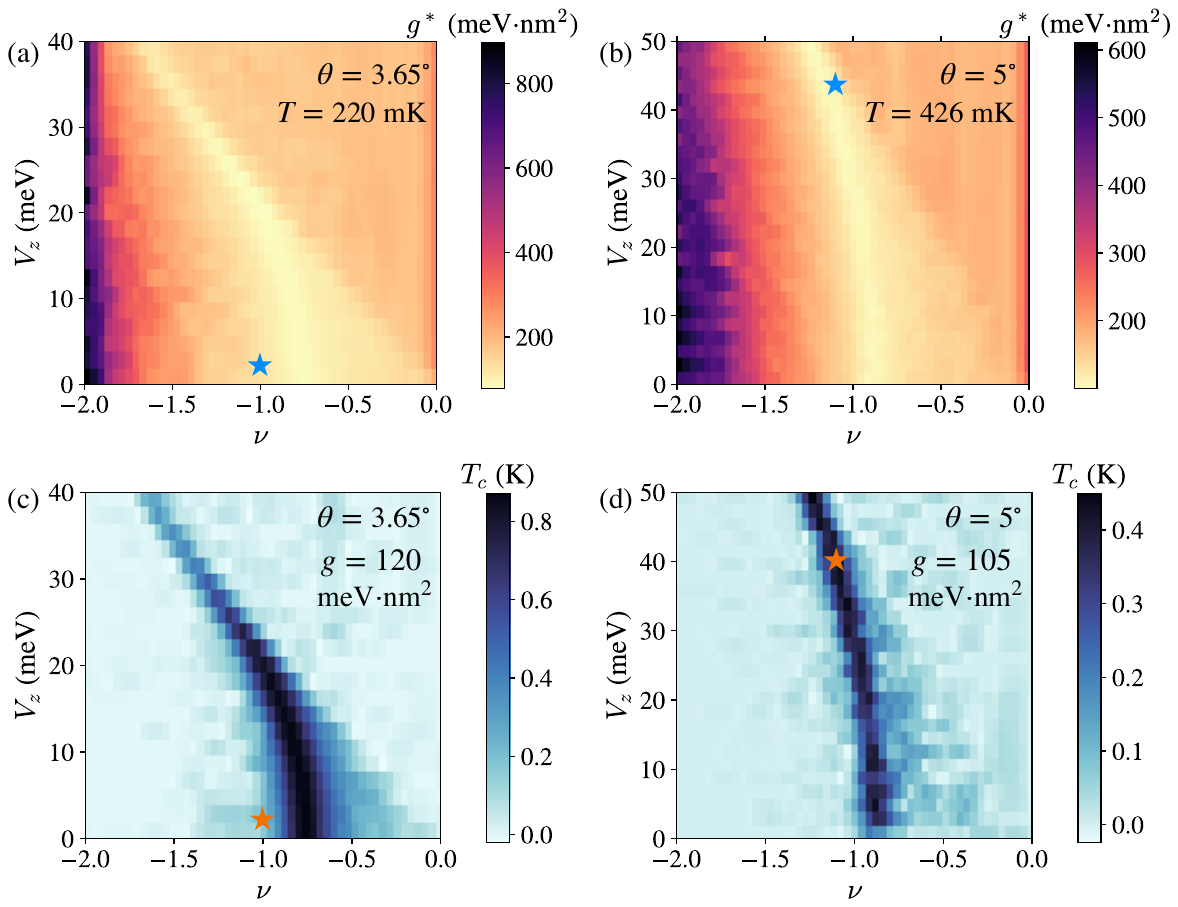}
\caption{\label{fig_gcrTc_map} {
(a-b) $g^*$ and (c-d) $T_c$ as a function of interlayer energy difference $V_z$ and filling factor $\nu$ for twist angles $\theta=3.65^\circ$ and $\theta=5^\circ$.
The experimental conditions realizing SC in Samples A and B are marked by stars.
Over the broad $(V_z, \nu)$ parameter range, $g^*$ values are similar in both samples with different twist angles.
  }}
\end{figure}

To further explore the superconducting properties, we show the phase diagrams of $g^*$ and $T_c$ across a broad range of displacement fields (or equivalently interlayer energy difference $V_z$) and filling factors for twist angles $\theta=3.65^\circ$ and $\theta=5^\circ$ in Fig.~\ref{fig_gcrTc_map}, in which the conditions for experimentally realized SC in Samples A and B are marked by stars.
In both cases, the minimum $g^*$ and maximum $T_c$ track the VHS. Remarkably, over the broad $(V_z, \nu)$ parameter space, $g^*$ values are similar in both samples (Fig.~\ref{fig_gcrTc_map}(a-b)) with different twist angles.
For the smaller twist angle (Fig.~\ref{fig_gcrTc_map}(a,c)), $T_c$ is maximized at $\nu \approx -0.8$ for a small displacement field, and at $\nu \lesssim -1$ for an intermediate displacement field, followed by a decrease in $T_c$ with increasing $V_z$.
For the larger twist angle (Fig.~\ref{fig_gcrTc_map}(b,d)), the VHS is pinned near $\nu=-1$, as well as maximum $T_c$. In Fig.~\ref{fig_gcrTc_map}(d), the maximum $T_c$ remains nearly constant with varying $V_z$ within the parameter range in our calculations. Note that the difference in maximum $T_c$ between Fig.~\ref{fig_gcrTc_map}(c) and (d) results from the different $g$ values used in these figures.

\section{C. Twist angle dependence}
\begin{figure}[!t]
\centering
\includegraphics[width=0.9\columnwidth]{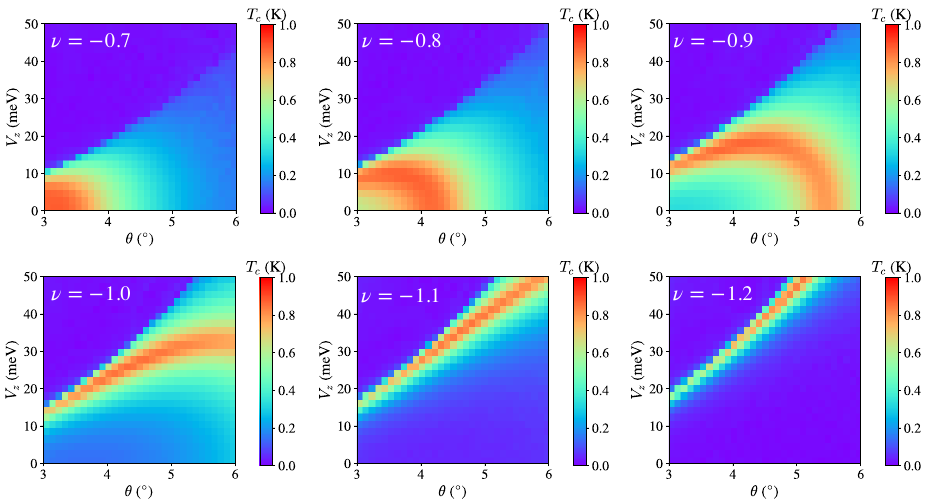}
\caption{\label{fig2_response} {
$T_c$ as a function of $V_z$ and twist angle $\theta$ for filling factors near half-filling, $\nu=-0.7$ to $-1.2$, using $g=120$ meV$\cdot$nm$^2$.
  }}
\end{figure}

\begin{figure}[!t]
\centering
\includegraphics[width=0.9\columnwidth]{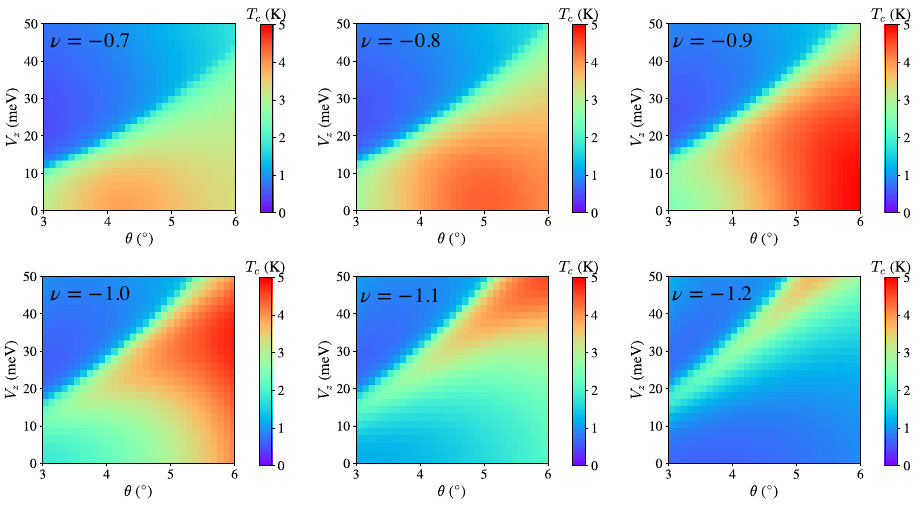}
\caption{\label{fig3_response} {
$T_c$ as a function of $V_z$ and twist angle $\theta$ for filling factors near half-filling, $\nu=-0.7$ to $-1.2$, using $g=200$ meV$\cdot$nm$^2$.
  }}
\end{figure}

In Figs. \ref{fig2_response} and \ref{fig3_response} we show additional phase diagrams illustrating $T_c$ as a function of both the displacement field ($V_z$) and the twist angle ($\theta$), for filling factors near half-filling. Using $g=120$ meV$\cdot$nm$^2$ (Fig.~\ref{fig2_response}), the maximum $T_c$ at $\nu=-1$ occurs at larger $V_z$ for larger twist angles. As the hole doping moves away from $\nu=-1$, the maximum $T_c$ shifts: for smaller hole doping, it moves to smaller twist angles and smaller $V_z$, whereas for larger hole doping, it shifts to larger twist angles and larger $V_z$.
Using $g=200$ meV$\cdot$nm$^2$ (Fig.~\ref{fig3_response}), we observe that the region of maximal $T_c$ expands over a larger range in the ($V_z$, $\theta$) phase space. Note that in the main text, we have estimated $g^* \sim 100$ to $120$ meV$\cdot$nm$^2$.

\end{document}